\documentclass[sigconf, anonymous=False]{acmart}
\usepackage{multirow}
\usepackage{tabularx} % For tabularx environment
\usepackage{lscape}
\usepackage{graphicx}
\usepackage{amsmath}
\usepackage{esvect}
\usepackage{caption}
\usepackage{comment}
\usepackage{tikz}
\usepackage{pgfplots}
\usepgfplotslibrary{groupplots}
\usepackage{caption}
\usepackage{subcaption}
\usepackage{float}
\restylefloat{table}
\AtBeginDocument{%
  \providecommand\BibTeX{{%
    \normalfont B\kern-0.5em{\scshape i\kern-0.25em b}\kern-0.8em\TeX}}}

\setcopyright{none}
\renewcommand\footnotetextcopyrightpermission[1]{}

\begin{document}

\title{Soft Prompt Tuning for Augmenting Dense Retrieval with Large Language Models}


\author{Zhiyuan Peng}
\authornote{Both authors contributed equally to this research.}
\affiliation{%
  \institution{Santa Clara University}
  \city{Santa Clara}
  \country{USA}}
\email{zpeng@scu.edu}

\author{Xuyang Wu}
\authornotemark[1]
\affiliation{%
  \institution{Santa Clara University}
  \city{Santa Clara}
  \country{USA}}
\email{xwu5@scu.edu}

\author{Qifan Wang}
\affiliation{%
  \institution{Meta AI}
  \city{Menlo Park}
  \country{USA}}
\email{wqfcr@meta.com}

\author{Yi Fang}
\authornote{Yi Fang is the corresponding author.}
\affiliation{%
  \institution{Santa Clara University}
  \city{Santa Clara}
  \country{USA}}
\email{yfang@scu.edu}

\renewcommand{\shortauthors}{Peng and Wu, et al.}

\begin{abstract}
Dense retrieval (DR) converts queries and documents into dense embeddings and measures the similarity between queries and documents in vector space. One of the major challenges in DR is the lack of domain-specific training data. While DR models can learn from large-scale public datasets like MS MARCO through transfer learning, evidence shows that not all DR models and domains can benefit from transfer learning. Recently, researchers have resorted to large language models (LLMs) to improve the zero-shot and few-shot DR models. However, the hard prompts or human-written prompts utilized in these works are suboptimal and the generated weak queries are often sensitive to the prompts. To tackle this, we propose soft prompt tuning for augmenting DR (SPTAR\footnote{\url{https://github.com/zhiyuanpeng/SPTAR.git}}): for each task, we leverage soft prompt-tuning to optimize a task-specific soft prompt on limited ground truth data and then prompt the LLMs to tag unlabeled documents with weak queries, yielding weak document-query pairs to train task-specific dense retrievers. We design a filter to select high-quality example document-query pairs in the prompt to further improve the quality of weak tagged queries. To the best of our knowledge, there is no prior work utilizing soft prompt tuning to augment DR models. Moreover, unlike much of the existing work, ours is based on popular open-source LLMs to ensure reproducible and deterministic results. Our experimental results demonstrate that SPTAR outperforms both unsupervised baselines and the recently proposed LLMs-based augmentation method for DR.
\end{abstract}
% evidence \cite{thakur2021beir} \cite{dai2022promptagator} shows that not all DR tasks and domains can benefit from transfer learning equally.




\settopmatter{printfolios=true}
\maketitle
\pagestyle{plain}
\section{Introduction}

Information retrieval (IR) plays a pivotal role in a wide array of applications, ranging from prominent web search engines such as Google and Bing to personalized recommendation systems like Walmart's product recommendations and Apple Music's song suggestions. Traditional IR methods, like TF-IDF and BM25 \cite{robertson2009probabilistic}, are built on token-level similarity matching, which can sometimes fall short due to a lexical gap \cite{berger2000bridging}. This gap occurs when semantically similar terms, such as synonyms, are overlooked because of their lexical differences. This oversight can potentially impact the quality of search results and the user experience. 

Given these constraints, researchers have turned to advancements in deep learning to tackle the lexical gap in conventional IR. One notable approach is Dense Retrieval (DR), which aims to capture the overarching semantic essence of content rather than fixating on individual tokens. DR models like dense passage retrieval (DPR) \cite{karpukhin2020dense} and ColBERT \cite{khattab2020colbert, santhanam2021colbertv2} encode each query or document into a dense vector, with the dimensionality determined by the neural networks. In practice, dense retrievers pre-compute document embeddings and construct an approximate nearest neighbor (ANN) index for rapid search. When a new query is introduced, only its embedding is computed and subsequently processed by the ANN search system. Unlike TF-IDF and BM25, DR places greater emphasis on assessing the similarity of the overall semantic context.

% Why use LLMs
While DR methods have made strides in bridging the lexical gap, they are still constrained by the limited availability of domain-specific training data, hindering their performance in specialized domains. Although some researchers have proposed to leverage transfer learning to mitigate this challenge, studies \cite{thakur2021beir, dai2022promptagator} indicate that not all DR models and domains can benefit from transfer learning equally. Recently, LLMs like CPT-3 \cite{brown2020language}, LLaMA \cite{touvron2023llama}, and Vicuna \cite{vicuna2023} have demonstrated potent zero-shot and few-shot learning. Rather than fine-tuning the LLMs on task-specific data, prompting integrates task instructions (e.g., TL;DR translate to English) and a few relevant examples as input and extracts the answers from the output of large language model (LLM). The terms ``hard promp'' and ``soft prompt'' refer to different approaches to guiding the LLM’s behavior during text generation or other tasks. A hard prompt \cite{DBLP:conf/nips/WenJKGGG23} involves using explicitly defined and unchangeable text inputs to instruct the model. The prompt does not involve additional training or fine-tuning of the model. On the other hand, a soft prompt involves using trainable vectors or learnable embeddings to guide the model’s behavior. Unlike hard prompts, soft prompts are not explicit text instructions but rather embeddings that influence the model’s output. These embeddings are typically learned through a process known as prompt tuning \cite{DBLP:conf/emnlp/LesterAC21,DBLP:conf/acl/LiL20,xprompt,PLP}. The existing work \cite{DBLP:conf/eacl/SchickS21, DBLP:conf/naacl/SchickS21} suggested that prompts provide a method for injecting task-specific guidance, which is beneficial in low-data regimes. Recent research \cite{scao2021many} further quantified this benefit through comprehensive testing of prompts. The results showed that well-crafted prompts can significantly reduce the dependency on large volumes of training data across downstream tasks. Both InPars \cite{DBLP:conf/sigir/BonifacioAFN22} and PROMPTAGATOR \cite{dai2022promptagator} employ hard prompts to guide LLMs in tagging unlabeled documents with weak queries, subsequently training task-specific retrievers. Nonetheless, hard prompts come with limitations: a) Crafting effective hard prompts is challenging and often requires iterative human effort, intuition, and sometimes a bit of luck; b) Even with hand-crafted prompts, the downstream tasks still underperform tuned models. For instance, compared with the performance of fine-tuned T5-XXL \cite{raffel2020exploring} on SuperGLUE \cite{wang2019superglue}, GPT-3 175B few-shot gets a 17.5 points smaller score despite using 16 times more parameters \cite{DBLP:conf/emnlp/LesterAC21}. These limitations of hard prompts underscore their effectiveness and addressing these challenges draws academic interest as well as generating industrial value.

% Soft prompt tuning in IR
Given the limitations of hard prompts, we investigate an alternative. Rather than utilizing humanly-readable words as hard prompts \cite{DBLP:conf/nips/PerezKC21}, the soft prompt \cite{DBLP:conf/emnlp/LesterAC21,DBLP:conf/acl/LiL20,xprompt,PLP} comprises a set of embeddings which are unrecognizable to humans and are prepended at the beginning of the neural network input. During the soft prompt tuning, the parameters of the LLM are frozen, and only the parameters associated with the soft prompt are updated. While both \cite{DBLP:conf/emnlp/LesterAC21} and \cite{DBLP:conf/acl/LiL20} demonstrate that soft prompts surpass the hard prompts, there is no work utilizing soft prompt tuning to augment DR. In this paper, we propose soft prompt tuning for augmenting DR (SPTAR). Specifically, for each task, we leverage soft prompt tuning to optimize the parameters associated with the soft prompt on limited ground truth data and then prompt the LLMs to tag unlabeled documents with weak queries, yielding enough weak document-query pairs to train task-specific retrievers. Moreover, we find that even with the optimized soft prompt, the quality of generated weak queries is sometimes sensitive to the example document-query pairs in the prompt. Thus, we designed a filter to select high-quality example document-query pairs in the prompt to further improve the quality of weakly tagged queries as well as the DR tasks. In addition, most of the existing work has been built on proprietary models hidden behind opaque API endpoints, which may produce non-reproducible or non-deterministic experimental results. Instead, our work is based on widely used open source LLMs \cite{DBLP:journals/corr/abs-2309-15088}. Our main contributions can be summarized as follows:
\begin{itemize}
\item To the best of our knowledge, our work stands as one of the early attempts of LLMs in combination with soft prompt tuning for enhancing DR tasks.
\item We introduce a soft prompt filter designed to curate document-query pairs within the prompt, thus enhancing the overall quality of the generated weak data. Additionally, we design a BM25 filter to reduce noise in the generated data, further improving performance.
\item We conduct a comprehensive set of experiments involving four datasets and seven retrievers and re-rankers, demonstrating the generality and superior performance of our approach over several state-of-the-art baselines.
\item Experiments are based on the recent open-source LLMs to ensure reproducible and deterministic experimental results. All code and data are publicly available\footnote{\url{https://github.com/zhiyuanpeng/SPTAR.git}}.
\end{itemize}
% Explore Prompt tuning in IR
\section{Related Work}
\subsection{Dense Retrieval}\label{sec: dense_retrieval}
DR converts the queries and documents into dense vectors on which the ANN index can be built for fast search. DPR \cite{karpukhin2020dense} employs a two-tower structure: one BERT model for queries and another for documents. For each query with one positive document and several negative documents, DPR measures the similarity between query embedding and document embeddings and then maximizes the log-likelihood of the positive passage. A variant of DPR is to utilize one BERT by concatenating query and document as input and extracting the query embedding and document embedding after the encoding. The query encoder and document encoder of ColBERT \cite{khattab2020colbert} \cite{santhanam2021colbertv2} share the same BERT but utilize a different special token following the ``[CLS]'' to distinguish query and document. Unlike DPR directly measures the similarity between query embedding and document embeddings, ColBERT introduces a late interaction mechanism. Specifically, for each token in the query, ColBERT computes its similarity with all the tokens in the document and applies a maximum pooling on these similarity scores. The similarity score of a pair of query and document is the summarization of all the scores after the maximum pooling. Given a query with one positive document and one negative document, ColBERT is optimized by the pairwise softmax cross-entropy loss over the computed scores of the positive and negative documents. ANCE \cite{xiong2020approximate} is a bi-encoder trained on (query, positive document, negative document) tuples where the negative document is retrieved from an ANN built on the checkpoint of the last step. TAS-B \cite{DBLP:conf/sigir/HofstatterLYLH21} groups queries by their embedding similarities and employs a training data sampling technique coupled with dual-teacher supervision distillation. Contriever \cite{DBLP:journals/tmlr/IzacardCHRBJG22} trains a bi-encoder model through contrastive learning. Instead of training the model on the labeled dataset, Contriever generates positive query-document pairs from unlabeled corpus by two strategies ``inverse cloze tasks'' and ``independent cropping''. ReContriever \cite{DBLP:conf/acl/LeiDCZYT23} adopts the same method as Contriever to generate the weak query-document pairs, but ReContriever scores the weak query-document pairs by itself during the training and the loss is weighted by the weights. BM25CE \cite{wang2020minilm} is a re-ranking-based DR. BM25CE first applies BM25 to retrieve documents and then employs the trained crossed-encoder to re-rank the retrieved documents. Our contribution is not to propose new dense retrievers but to propose a novel method to augment the existing dense retrievers.

\subsection{Data Augmentation for Dense Retrieval} \label{sec: DAforDR}

For DR datasets, usually, only a fraction of documents are labeled with queries, for instance, MS MARCO \cite{nguyen2016ms}, a widely used dataset in DR, has a corpus of 8841823 documents but only has 532761 training document-query pairs. Given DR demands substantial training data to achieve quality dense embeddings, some researchers have turned to data augmentation to generate more document-query pairs to train better dense embeddings. InPars \cite{DBLP:conf/sigir/BonifacioAFN22} feeds a task-specific human-written prompt and 3 example document-query pairs to a 6B GPT-3 \cite{brown2020language} model Curie to generate 100K weak document-query pairs and selects the top 10K queries with respect to the probability of query $q$ to augment the training data. InPars \cite{DBLP:conf/sigir/BonifacioAFN22} employs the same dense retrieval model proposed in \cite{nogueira2020document}, which treats the retrieval as a sequence-to-sequence task by concatenating a query and a document as input to T5 mode and outputs the relevance score. Improved variations of InPars \cite{DBLP:conf/sigir/BonifacioAFN22}, such as InPars-v2 \cite{DBLP:journals/corr/abs-2301-01820} and InPars-Light \cite{DBLP:journals/corr/abs-2301-02998}, have been introduced to enhance the original methodology. Like InPars \cite{DBLP:conf/sigir/BonifacioAFN22}, PROMPTAGATOR \cite{dai2022promptagator} also feeds a task-specific human-written prompt and at most 8 example document-query pairs to LLM to generate weak data. Instead of selecting the top weak queries by their probabilities, PROMPTAGATOR first trains a filter on uncleaned document-query pairs to filter the weak queries by dropping the weak queries that cannot retrieve their paired documents in the Top-$k$ retrieved documents. By repeating this process multiple times, the filter significantly improves the performance of a dual-encoder DPR retriever. Besides, PROMPTAGATOR \cite{dai2022promptagator} utilizes a much bigger LLM: a 175B model Flan \cite{wei2021finetuned} which cannot be accessed by most researchers. DAR \cite{DBLP:conf/acl/JeongBCHP22} argues that the method that generates queries from unlabeled documents is costly as well as does not add variations to the documents. To do data augmentation efficiently, DAR \cite{DBLP:conf/acl/JeongBCHP22} not only interpolates two different document representations associated with the
labeled query but also stochastically perturbs the representations of labeled documents in embedding space. RocketQA \cite{DBLP:conf/naacl/QuDLLRZDWW21} applies a pre-trained cross-encoder retriever to retrieve positive and negative documents for a new collection of queries with high confidence scores. RocketQAv2 \cite{DBLP:conf/emnlp/RenQLZSWWW21} augments the DR by jointly optimizing the bi-encoder structure DR and cross-encoder structure reranking model to have similar output distributions. DRAGON \cite{DBLP:journals/corr/abs-2302-07452} fuses multiple teacher models by progressively training the base DR model.

\subsection{LLMs in Dense Retrieval}\label{IRwithLLMs}

Most of the current literature in this domain explores the potential of LLMs to improve DR tasks through various data generation techniques, including query generation \cite{DBLP:conf/sigir/BonifacioAFN22, DBLP:journals/corr/abs-2301-01820, DBLP:journals/corr/abs-2301-02998, dai2022promptagator, sachan2022improving, DBLP:journals/corr/abs-2212-10496}, relevance generation \cite{DBLP:journals/corr/abs-2211-09110}, and permutation generation \cite{ma2023zero, sun2023chatgpt, DBLP:journals/corr/abs-2306-17563}. PROMPTAGATOR \cite{dai2022promptagator} and InPars \cite{DBLP:conf/sigir/BonifacioAFN22} with its variations InPars-v2 \cite{DBLP:journals/corr/abs-2301-01820} and InPars-Light \cite{DBLP:journals/corr/abs-2301-02998} are illustrated in section \ref{sec: DAforDR}. UPR \cite{sachan2022improving} utilizes LLM as a zero-shot reranker to re-rank the passages retrieved by retrievers like BM25 and DPR. Given a query, for each retrieved passage, UPR utilizes a prompt ``\textit{Please write a question based on this passage}'' to prompt a LLM and computes the average log-likelihood of the question tokens conditioned on the input document as the relevance score. Due to the intensive computational resources required to train LLMs, all these works utilize LLMs as query generators instead of fine-tuning them. HyDE \cite{DBLP:journals/corr/abs-2212-10496} leverages LLMs to augment queries by generating hypothetical documents, effectively capturing relevance patterns for unsupervised retrieval. LRL \cite{ma2023zero} trains a listwise zero-shot re-ranker that leverages LLMs without task-specific supervised training. Unlike pointwise re-rankers, LRL considers all candidate documents to determine their relative ranking positions. Another approach involves instructional permutation generation \cite{sun2023chatgpt}, where the focus is on instructing LLMs to directly output permutations of passages. Permutation distillation techniques are employed to transfer the passage ranking capabilities of ChatGPT into a smaller, specialized ranking model. While these works utilize LLMs as query generators without fine-tuning, our SPTAR approach takes a different approach. We first perform soft prompt tuning to optimize task-specific soft prompts and then employ data filtering to enhance the quality of the generated weak data.

\subsection{Prompt Tuning}

Prompt tuning offers a promising avenue for adapting pre-trained LLMs to specific tasks by focusing on tuning the prompt module instead of fine-tuning the entire model \cite{tam2022parameter}. Prefix-Tuning \cite{DBLP:conf/acl/LiL20} introduces a prompt module with learnable parameters $\theta$ outputting embeddings which are prepended to the embeddings of other inputted tokens. This approach preserves the original training objective intact while updating only the prefix parameters $\theta$ through gradient descent for each task. Another similar technique, referred to as ``\textit{gisting}'' \cite{DBLP:journals/corr/abs-2304-08467}, compresses arbitrary prompts into a condensed set of virtual ``\textit{gist}'' tokens using a meta-learning approach. Building upon T5 \cite{raffel2020exploring}, \citet{DBLP:conf/emnlp/LesterAC21} propose a method where the learnable embeddings of a task-specific prompt are prepended to the encoder's output. The concatenated embeddings are then passed through the decoder to compute the training objective. This approach enables the model to incorporate task-specific information into the decoding process. \citet{zhou2022dual} introduce Dual Context-guided Continuous Prompt (DCCP), which employs soft prompt tuning using dual inputs: context-aware prompt and label-aware context representations. This approach leverages both prompt information and contextual understanding to enhance the model's performance. Prompt tuning can benefit from multi-task learning. For instance, ATTEMPT proposed by \citet{DBLP:conf/iclr/WangPKF0K23} introduces a multi-task tuning method that transfers knowledge across different tasks through a mixture of soft prompts. In the context of Multilingual Information Retrieval, \citet{DBLP:journals/corr/abs-2305-09025} explore a soft prompt decoding approach that treats retrieval in each language as a separate task while jointly modeling them to capture shared underlying structures. They use decomposable prompts in KD-SPD to model languages, highlighting that languages share common features and concepts despite their unique properties. Regarding IR tasks, DPTDR by \citet{tang2022dptdr} employs a dual-encoder, two RoBERTa models, for retrieval. It initializes the dual-encoder through contrastive learning and appends learnable soft prompts for query and document. Both the dual-encoder and the learnable prompts are updated during the training process. 

In contrast, unlike the current DR data augmentation works that only prompt LLMs to generate weak queries for unlabeled documents with a few labeled document-query pairs as examples, we propose to learn task-specific soft prompts on a small proportion of the labeled data and a novel soft prompt filter method to select high-quality example document-query pairs in the prompt to improve the DR tasks further. The whole augmentation pipeline makes our approach different from the current works.

\section{Soft Prompt Tuning for Augmenting Dense Retrieval} \label{sec: SPTAR}

\begin{table*}[t!]
\centering
  % \begin{center}
  % \resizebox{\linewidth}{!}{%
    % \begin{tabular}{p{0.095\textwidth}|p{0.74\textwidth}} % <-- Alignments: 1st column left, 2nd middle and 3rd right, with vertical lines in between
    % \renewcommand{\arraystretch}{1.2}
    \begin{tabular}{p{0.18\textwidth}|p{0.70\textwidth}}
    % \begin{tabular}
      \toprule
      \bf Notation & \bf Definition \\
      \hline
      $q$, $d$ & query, document\\
      ${D}_{train}$& collection of query-document pairs for training \\ 
        ${D}_{test}$& collection of query-document pairs for testing \\
      ${D}_{eval}$& collection of query-document pairs for evaluation \\
      $C$ & collection of all the documents \\
      $C_{unlabeled}$ & unlabeled documents in $C$ \\
      $\Phi$ & freezed original parameters of large language model \\
      $S_{train}^X$ & collection of $X$ sampled queries associated with corresponding documents from $D_{train}$ \\
      $f_{\theta}(\cdot)$ & prompt's embedding layer parametrized by $\theta$ \\
      $f_{\theta}(s)$ & soft prompt initialized based on a hard prompt $s$ with a length of $l_s$ \\
      $(d_{m}, q_{m})_{m=1}^M$ & collection of $M$ sampled document-query pairs from $S_{train}^X$, as examples in prompt \\
      $c_j$ & concatenation of $(d_j, q_j)\in (d_{j}, q_{j})_{j=1}^{NumPair(X)-M}$ and all the example pairs $(d_{m}, q_{m})_{m=1}^M$ \\
      $(d_{j}, q_{j})_{j=1}^{NumPair(X)-M}$ & for each epoch, loss is computed on $NumPair(X)-M$ document-query pairs from $S_{train}^X$\\
      $p_{\theta, \Phi}(q_{j}|t_{j})$ & the probability of $q_{j}$ conditioned on $t_{j}$ given parameters $\theta$ and $\Phi$\\
      $h_{j,i}$ & output hidden vector of $i^{th}$ time step for $j^{th}$ instance\\
      $z_{j,i}$ & ID of $i^{th}$ token of $j^{th}$ instance\\
      $L$ & loss function \\
      $S_{eval}^Y$ & collection of $Y$ queries associated with corresponding documents from $D_{eval}$ \\
      $W_{large}$ & collection of 100K sampled documents from $C_{unlabeled}$ and each document has a generated weak query\\
      $W_{small}$ & collection of 5000 sampled documents $W_{large}$ \\
      $F_{k}(\cdot)$ & top $k$ weak data filter function \\
      $F_{k}(W_{large})$ & $W_{large}$ filtered by weak data filter $F_{k}$ \\
      \bottomrule
    \end{tabular}
    \caption{Summary of notation.}
    \label{tab: notation}
  % \end{center}
\end{table*}

\begin{figure}[t!]
\centering
\includegraphics[width=0.7\linewidth]{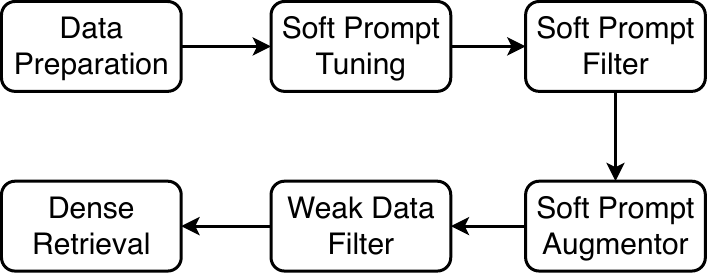}
\caption{The pipeline of the proposed Soft Prompt Tuning for Augmenting dense Retrieval (SPTAR).}
\label{fig: workflow} 
\end{figure}

As shown in Figure \ref{fig: workflow}, SPTAR comprises six modules: a) data preparation; b) soft prompt tuning; c) soft prompt filter; d) soft prompt augmentor; e) weak data filter; f) DR. In Section \ref{sec: data_preparation}, we elaborate on how to generate the training and evaluation datasets of soft prompt tuning. With the training and evaluation datasets, we conduct soft prompt tuning (Section \ref{sec: soft_prompt_tuning}) to learn a task-specific soft prompt. To further improve the quality of the weak generated queries, we introduce the soft prompt filter (Section \ref{sec: soft_prompt_filter}) which identifies optimal example document-query pairs to optimize the task-specific prompt. We then prompt LLMs to generate weak queries for unlabeled documents (Section \ref{sec: soft_prompt_augmentor}), yielding enough training data to train DR. Finally, we train the DR (Section \ref{sec: SPTAR_dr}) models on filtered weak data (Section \ref{sec: weak_data_filter}). The notations used in this paper are provided in Table \ref{tab: notation}.

\subsection{Data Preparation}
\label{sec: data_preparation}
We study the augmentation of DR using limited data. The initial step involves sampling a small dataset on which we fine-tune a task-specific soft prompt. We define dataset $D$ as $D = \{(q_n, d_n)\}_{n=1}^N$ where for each query $q_n$, there is a relevant document $d_n$. There may exist duplicated queries as one query may have multiple relevant documents. This domain-specific dataset $D$ is categorized into train, test, and evaluation subsets, denoted as $D_{train}$, $D_{test}$, and $D_{eval}$, respectively. Apart from dataset $D$, there is a much bigger document collection $C$ which contains all the documents in $D$ but has more unlabeled documents denoted as $C_{unlabeled}$. After training, DR encodes all the documents in $C$ into vectors. When a new query comes in, DR encodes the query into a vector and searches the top-$k$ similar documents in vector space.

\begin{figure*}[t!]
\centering
\includegraphics[width=1.0\linewidth]{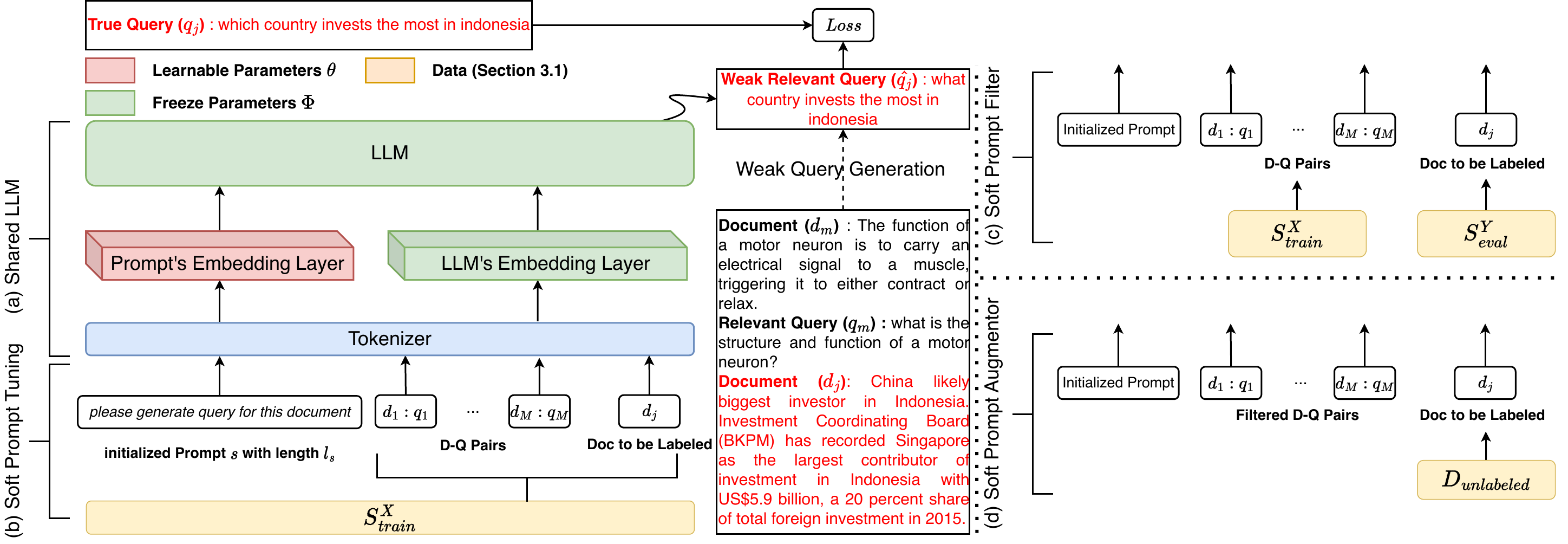}
\caption{The main architecture of the proposed SPTAR: a) The same LLM is shared by soft prompt tuning module, soft prompt filter module and soft prompt augmentor module; b) soft prompt tuning module fixs the LLM's original parameters $\Phi$ and only fine-tune the parameters of soft prompt's embedding layer $\theta$ on the sampled small dataset (Section \ref{sec: data_preparation}); c) soft prompt filter module fixs the learned parameters $\theta^{\ast}$, and for each group of sampled example document-query pairs, computes the loss on evaluation dataset. The group of example document-query pairs with the smallest loss will be utilized in the soft prompt augmentor module; d) with the learned parameters $\theta^{\ast}$ and a group of filtered example document-query pairs, the soft prompt augmentor module iterates over the unlabeled document dataset $D_{unlabeled}$ to generate weak queries.}
\label{fig: SPTAR} 
\end{figure*}

We randomly sample document-query pairs from the original training dataset $D_{train}$ to construct the training and evaluation datasets for the soft prompt module, namely $S_{train}^{X}$ and $S_{eval}^{Y}$ where indices X and Y signify the number of distinct queries within the training and evaluation datasets respectively. The function $NumPair(x)$ designates the quantity of document-query pairs given $x$ distinct queries in the dataset. Since each query may have more than one positive document, $NumPair(x)$ may be bigger than $|x|$. Hence, $S_{train}^{X}$ contains $NumPair(X)$ document-query pairs, similarly, $S_{eval}^{Y}$ comprises $NumPair(Y)$ document-query pairs. For illustration, in our experiment, we draw 50 unique queries and their corresponding documents from the training dataset $S_{train}$ to form $S_{train}^{50} (X=50)$. From the remaining data in $S_{train}$, we randomly select 100 unique queries and their associated documents to compose $S_{eval}^{100} (Y=100)$. $S_{train}^{50}$ serves for optimizing the soft prompt, while $S_{eval}^{100}$ is employed to assess the model's convergence, enabling us to terminate the training process in advance and mitigate overfitting risks. We also tried other values of $X$, and the influence of $X$ is studied in Section \ref{sec: train_num_prompt}. 

\subsection{Soft Prompt Tuning} \label{sec: soft_prompt_tuning}
Soft prompts \cite{MixPAVE,E2VPT} introduce a novel technique to steer a model's behavior without the need for extensive fine-tuning. Unlike hard prompts, which are human-readable instructions, soft prompts comprise trained embeddings optimized for specific tasks. The soft prompt tuning module learns a task-specific soft prompt on a small proportion of labeled data. Figure \ref{fig: SPTAR} (b) illustrates the structure of the soft prompt tuning module, where the red boxes represent the parameters $\theta$ to be optimized during model training and the green boxes represent LLM's original parameters $\Phi$ that are retained during the training. $s$ represents the initialized hard prompt with size $l_{s}$, like repeating "please generate query for document" until the length of $s$ equals $l_{s}$. Let $f_{\theta}(\cdot)$ denote the prompt's embedding layer implemented by an embedding matrix initialized as the embeddings of $s$ encoded by LLM's original embedding layer. $f_{\theta}(s)$ represents the soft prompt.

For each training epoch, we first randomly sample $M$ document-query pairs from training dataset $S_{train}^{X}$ as example document-query pairs $(d_{m}, q_{m})_{m=1}^M$, then iterate over the left document-query pairs $(d_{j}, q_{j})_{j=1}^{NumPair(X)-M}$ to compute loss. Example pairs $(d_{m}, q_{m})_{m=1}^M$ are concatenated with each pair $(d_{j}, q_{j})$ by keywords like ``\textit{document}'' and ``\textit{query}'' as $c_{j}$. Finally, we concatenate $s$ with $c_{j}$ as one training instance $t_{j} = [s; c_{j}]$ and there are $NumPair(X)-M$ instances in each epoch. When $t_j$ is inputted into the soft prompt tuning module, it is first tokenized into a list of IDs $z_{j}$ indexed by $i$ then the embeddings of IDs are extracted and fed into the following layers to compute the hidden vectors. $f_{\theta}(\cdot)$ takes the IDs of $s$ as inputs and outputs its embeddings while the embeddings of $c_{j}$ are generated by LLM's original embedding layer. For simplicity, we postulate that each token in $t_{j}$ has one corresponding ID in $z_{j}$. For training instance $t_{j}$, the hidden vector of $i^{th}$ time step is defined as $h_{j,i} \in \mathbb{R}^d$ where $h_{j,i}=\left[h_{j,i}^{(1)} ; \cdots ; h_{j,i}^{(k)}\right]$ and $k$ is the number of layers in LLM. The objective function for training is given by:
\begin{equation}
\max _\theta \log p_{\theta,\phi}(q_{j} \mid t_{j})=\max _\theta \sum_{i\in idx_{q_{j}}} \log p_{\theta,\phi}\left(z_{j,i} \mid h_{j,<i}\right)
\end{equation}
where $idx_{q_j}$ denotes the indexes corresponding to the IDs of $d_j$. Additionally, $p_{\theta,\phi}\left(z_{j,i} \mid h_{j,<i}\right)$ signifies the probability of the subsequent token with ID $z_{j,i}$. For loss function $L$, we employ the negative log-likelihood defined as:
\begin{equation} \label{eq: loss}
L = - \log p_{\theta,\phi}(q_{j} \mid t_{j})
\end{equation}
We implemented our soft prompt tuning module based on a public prompt tuning package PEFT \cite{peft}. 

\subsection{Soft Prompt Filter}
\label{sec: soft_prompt_filter}

During the development of the soft prompt module, we observe that the choice of example document-query pairs $(d_{m}, q_{m})_{m=1}^M$ profoundly affects the quality of text generation. Therefore, upon completing the soft prompt training, with the learned parameters $\theta^{\ast}$, we try to select the best group of document-query pairs from $S_{train}^{X}$ as example document-query pairs in soft prompt augmentor. For $M=2$, there are 1225 ($50*49/2$) groups of example pairs, which makes it impractical to evaluate all. To reduce the computation complexity, we randomly sample $X$ groups of example pairs from $S_{train}^{X}$ to evaluate them on the evaluation dataset $S_{eval}^{Y}$ and the group of example pairs with the best evaluation metric will be chosen as the example pairs in soft prompt augmentor. As shown in Figure \ref{fig: SPTAR} (c), the only difference between soft prompt tuning and soft prompt filter is the dataset where the $d_{j}$ comes from. Suppose we sampled $X$ groups of document-query pairs each of which has $M$ document-query pairs $(d_{m}, q_{m})_{m=1}^M$. Evaluation dataset $S_{eval}^{Y}$ has $Num(Y)$ document-query pairs and example pairs $(d_{m}, q_{m})_{m=1}^M$ are concatenated with each pair $(d_{j}, q_{j})$ by keywords like ``\textit{document}'' and ``\textit{query}'' as $c_{j}$.  Then, $c_{j}$ is concatenated with the initialized prompt $s$ as $t_{j}=[s, c_{j}]$. The evaluation metric is the same as the loss function $L$ (Equation \ref{eq: loss}). 
We study the effectiveness of soft prompt filter in Section \ref{sec: prompt_filter} and the filtered example document-query pairs are documented in Appendix \ref{sec: example_pairs_ms} and \ref{sec: example_pairs_fiqa}.

\subsection{Weak Data Filter} 
\label{sec: weak_data_filter}

Both InPars \cite{DBLP:conf/sigir/BonifacioAFN22} and PROPAGATE \cite{dai2022promptagator} emphasize the importance of filtering weak document-query pairs as the generated weak queries are not guaranteed to be always relevant to the input documents. Employing the methodology from InPars \cite{DBLP:conf/sigir/BonifacioAFN22}, we clean the weak data. Upon acquiring these generated weak pairs (Section \ref{sec: soft_prompt_augmentor}), we apply a BM25-based filtering: For each weak query, BM25 retrieves the top $k$ documents from the corpus $C$. If the document linked to a weak query isn’t among the top $k$ results, the pair gets discarded. This filtering approach is denoted as $F_{k}(\cdot)$. For datasets MS MARCO and FiQA-2018, we experimented with different top $k$ values from the set $\{10,30,50,70\}$ and reported the best results. The effectiveness of this weak data filter module is discussed in Section \ref{sec: ab_data_filter}.

\subsection{Soft Prompt Augmentor} 
\label{sec: soft_prompt_augmentor}

Generating high-quality queries for unlabeled documents remains a formidable challenge. Our soft prompt augmentor module harnesses both the potency of the learned soft prompts and the context offered by the best example document-query pairs, providing a synergistic effect that ensures not just relevance, but also the superior quality of the generated queries. As shown in Figure \ref{fig: SPTAR} (d), with the learned parameters $\theta^{\ast}$ and the filtered group of example document-query pairs, soft prompt augmentor generates a weak query for an unlabeled document $d_{j}$ sampled from $D_{unlabeled}$. In this paper, for each dataset, we first created two weak datasets: a) $W_{large}$. $100K$ unlabled documents are sampled from $D_{unlabeled}$ to generate their weak queries. If the number of unlabeled documents in $D_{unlabeled}$ is smaller than $100K$, all the unlabeled documents are utilized to generate weak queries; b) $W_{small}$. 5000 document-query pairs are sampled from $W_{large}$. Then, we filtered $W_{large}$ and $W_{small}$ by weak data filter, described in Section \ref{sec: weak_data_filter}, to get $F_{k}(W_{large})$ and $F_{k}(W_{small})$. During the weak query generation process, LLM not only utilizes the soft prompt embeddings to capture domain-specific information but also benefits from the supplementary context provided by the best example document-query pairs.
%This combination of information enhances the quality of the resulting weak queries, making them more relevant and effective for downstream tasks.

\subsection{Dense Retrieval}
\label{sec: SPTAR_dr}

DR serves as the concluding step in our methodology, where we harness the capabilities of neural networks to retrieve relevant documents. We conducted the experiments on five popular dense retrievers: DPR, ColBERT, TAS-B, Contriever, and ReContriever. The descriptions of the models can be found in Section \ref{sec: dense_retrieval}. For TAS-B, we used pre-trained bi-encoder model \footnote{https://huggingface.co/sentence-transformers/msmarco-bert-base-dot-v5} for clustering queries and cross-encoder model \footnote{https://huggingface.co/cross-encoder/ms-marco-MiniLM-L-6-v2} and ColBERTv2 \footnote{https://downloads.cs.stanford.edu/nlp/data/colbert/colbertv2/colbertv2.0.tar.gz} for teacher models. For Contriever, we refer to the officially released checkpoint \footnote{https://huggingface.co/facebook/contriever} as $Contriever_{base}$ and used it for initial evaluations. Additionally, we fine-tuned $Contriever_{base}$ as a bi-encoder dense retrieval model, denoted $Contriever_{be}$. Similary, for ReContriever, we have $ReContriever_{base}$ \footnote{https://huggingface.co/Yibin-Lei/ReContriever} and $ReContriever_{be}$. Further, we incorporated the cross-encoder model BM25CE, as established in previous literature \cite{wang2020minilm}, which re-ranks the top 1000 items retrieved by BM25. Like BM25CE, we employed the DPR model as a bi-encoder to re-rank the top 1000 items retrieved by BM25, referring to this method as BM25BE.

\begin{table*}[t!]
\centering
\resizebox{\linewidth}{!}{%
\begin{tabular}{c|ccccccccc}
\toprule
\multirow{2.5}{*}{\textbf{Task}} & \multirow{2.5}{*}{\textbf{Domain}} & \multirow{2.5}{*}{\textbf{Dataset}} & \multicolumn{1}{c}{\textbf{Train}} & \multicolumn{1}{c}{\textbf{Eval}} & \multicolumn{3}{c}{\textbf{Test}} & \multicolumn{2}{c}{\textbf{Avg. Word Lengths}}\\
\cmidrule(lr){4-4} \cmidrule(lr){5-5} \cmidrule(lr){6-8} \cmidrule(lr){9-10}
 &  &  & \textbf{\#Pairs} & \textbf{\#Query} & \textbf{\#Query} & \textbf{\#Corpus} & \textbf{Avg. D/Q} & \textbf{Query} & \textbf{Document}\\
\midrule
Passage Retrieval & Misc. & MS MARCO \cite{nguyen2016ms} & 532,761 & N/A & 6,980 & 8,841,823 & 1.1 & 5.96 & 55.98 \\
Passage Retrieval & Misc. & DL2019 \cite{DBLP:journals/corr/abs-2003-07820} & 532,761 & N/A & 43 & 8,841,823 & 215.3 & 5.96 & 55.98 \\
Passage Retrieval & Misc. & DL2020 \cite{DBLP:conf/trec/CraswellMMYC20} & 532,761 & N/A & 54 & 8,841,823 & 210.9 & 5.96 & 55.98 \\
Question Answering & Finance & FiQA-2018 \cite{maia201818} & 14,166 & 500 & 648 & 57,638 & 2.6 & 10.77 & 132.32\\
\bottomrule
\end{tabular}}
\caption{Statistics of datasets in BEIR benchmark. Avg. D/Q indicates the average number of relevant documents per query.}
\label{tab: dataset}
\end{table*}

\section{Experimental Setup}
\subsection{Datasets}\label{sec: dataset}

\begin{table}[t]
\centering
\renewcommand{\arraystretch}{1.2}
\begin{tabular}{c|ccc}
    \toprule
    Model & Train & Eval & Test\\
    \midrule
    BM25           &         N/A                                         &      N/A            & $D_{test}$\\
    $Contriever_{base}$&N/A&N/A&$D_{test}$\\
    $ReContriever_{base}$&N/A &      N/A            & $D_{test}$\\
    W/O Aug        & $S_{train}^{50}$ + $S_{eval}^{100}$                & $D_{eval}$         & $D_{test}$\\
    InPars        & $S_{train}^{50}$ + $S_{eval}^{100}$ + $F_{k}(W_{large})$ & $D_{eval}$         & $D_{test}$\\
    SPTAR-Tuning  & $S_{train}^{50}$                                   & $S_{eval}^{100}$   & N/A         \\
    SPTAR-DR       & $S_{train}^{50}$ + $S_{eval}^{100}$ + $F_{k}(W_{large})$ & $D_{eval}$         & $D_{test}$\\
    \bottomrule
\end{tabular}
\caption{Dataset partition for different methods.}
\label{tab: datasets-partitioin}
\end{table}
\renewcommand{\arraystretch}{1.0}

Experiments were performed on four datasets MS MARCO \cite{nguyen2016ms} and FiQA-2018 \cite{maia201818}, sourced from BEIR \cite{thakur2021beir} and DL2019 \cite{DBLP:journals/corr/abs-2003-07820}, DL2020 \cite{DBLP:conf/trec/CraswellMMYC20}. The description of the four datasets can be found in Table \ref{tab: dataset}. We follow BEIR \cite{thakur2021beir} to report the metrics on the evaluation dataset instead of test data for MS MARCO, so, for MS MARCO, $D_{test}$ is the same as $D_{eval}$. 

As shown in Table \ref{tab: datasets-partitioin}: a) BM25, $Contriever_{base}$ and $ReContriever_{base}$ are evaluated on the original testing split $D_{test}$; b) W/O Aug models are trained on datasets $S_{train}^{50}$ and $S_{eval}^{100}$ utilized to fine-tune the soft prompt; c) InPars \cite{DBLP:conf/sigir/BonifacioAFN22} models are trained on $S_{train}^{50}$ and $S_{eval}^{100}$ plus $F_{k}(W_{large})$ ($W_{large}$ filtered by $F_{k}$, Section \ref{sec: weak_data_filter}) generated by human-written prompts. d) SPTAR's soft prompt tuning module (SPTAR-Tuning) is trained on $S_{train}^{50}$ and evaluated on $S_{eval}^{100}$; SPTAR's DR models (SPTAR-DR) are trained on $S_{train}^{50}$ and $S_{eval}^{100}$ plus $F_{k}(W_{large})$ ($W_{large}$ filtered by $F_{k}$, Section \ref{sec: weak_data_filter}) generated by soft prompt augmentor (Section \ref{sec: soft_prompt_augmentor}); e) W/O Aug, InPars \cite{DBLP:conf/sigir/BonifacioAFN22} and SPTAR are all evaluated and tested on the same splits for a fair comparison; f) For $F_{k}(\cdot)$, we tried $k\in(10,30,50,70)$ and selected the $k$ with the best NDCG@10 score on evaluation dataset.

\subsection{Training Details}

To train the soft prompt module, we performed fine-tuning using two open-source LLMs: LLaMA-7B and Vicuna-7B. The specific training hyper-parameters are documented in Table \ref{tab: hyper-tuning}. 
\begin{table}[H]
\centering
\begin{tabular}{c|cc}
    \toprule
    Hyperparameters & LLaMA-7B & Vicuna-7B\\
    \midrule
    Batch Size & 4 & 2\\
    Max Length & 1024 & 1024\\
    Learning Rate & $3e-2$ & $3e-2$\\
    Optimizer & AdamW & AdamW\\
    Early Stop & 5 & 5\\
    Max epochs & 100 & 100\\
    GPU & 1 A100 (80G) & 1 A100 (80G)\\
    \bottomrule
\end{tabular}
\caption{Hyperparameters of soft prompt tuning}
\label{tab: hyper-tuning}
\end{table}

\begin{table*}[t!]
\centering
\resizebox{\linewidth}{!}{%
\begin{tabular}{c|cccccc}
    \toprule
    Hyperparameters & DPR & ColBERT & $Contriever_{be}$ &$ReContriever_{be}$ & BM25CE & TAS-B\\
    \midrule
    Batch Size & 32 & 32 & 32 & 32 & 96 & 40\\
    Max Length & 350 & 350 & 350 & 350 & 512 & 300\\
    Learning Rate & $2e-5$ & $2e-5$ & $2e-5$ & $2e-5$ & $2e-5$ & $2e-5$\\
    DDP & No & Yes & No & No & No & No\\
    Optimizer & AdamW & AdamW & AdamW & AdamW & AdamW & AdamW\\
    Early Stop & 10 & None & 10 & 10 & 10 & 10\\
    Max epochs & 20 & 20 & 20 & 20 & 20 & 20\\
    GPU & 4 A100s (40G) & 4 A100s (40G) & 4 A100s (40G) & 4 A100s (40G) & 4 A100s (40G) & 4 A100s (40G)\\
    \bottomrule
\end{tabular}}
\caption{Hyperparameters of DR Models}
\label{tab: hyper-dr}
\end{table*}

The training hyper-parameters of dense retrievers are in Table \ref{tab: hyper-dr}. For ColBERT, there is no early stop in the official code and we saved a checkpoint after each epoch. After training, we manually evaluated some checkpoints (3, 5, 10, 15, 18, 20) and reported the testing results of the checkpoint with the highest NDCG@10 score.

\subsection{Evaluation Metrics}
In the context of text generation models, Perplexity is a commonly employed metric that quantifies the level of uncertainty exhibited by a language model when generating new tokens. This metric is defined as the exponentiated average negative log-likelihood of a sequence, and a lower perplexity value indicates a higher-quality language model. Perplexity is used to evaluate the soft prompt tuning and soft prompt filter modules.

In our evaluation of DR models, we adhere to the metrics established in previous studies \cite{DBLP:conf/sigir/LinMLYPN21, kamalloo2024resources}. For the MS MARCO and FiQA-2018 datasets, we utilize Mean Reciprocal Rank at 10 (MRR@10) and Recall@100. For the DL2019 and DL2020 datasets, we employ Mean Average Precision (MAP), Normalized Discounted Cumulative Gain at 10 (nDCG@10), and Recall@100. For the BM25CE and BM25BE models, which both re-rank the top 1000 items retrieved by BM25, Recall@100 is used to specifically evaluate their re-ranking effectiveness. These metrics together provide a comprehensive assessment of how augmented queries influence the performance of DR models.

\subsection{Baseline Methods}

Our study incorporates five baseline methods: BM25, $Contriever_{base}$, $ReContriever_{base}$, Without Augmentation (W/O Aug), and InPars \cite{DBLP:conf/sigir/BonifacioAFN22} (Section \ref{IRwithLLMs}). The training, evaluation, and testing datasets are documented in Section \ref{sec: dataset}. For BM25 \cite{robertson2009probabilistic}, we use Anserini \cite{lin2016toward} with the default Lucene parameters ($k=0.9$ and $b=0.4$). TAS-B, $Contriever$, and $ReContriever$ are described in Section \ref{sec: SPTAR_dr}. The differences between InPars \cite{DBLP:conf/sigir/BonifacioAFN22} and SPTAR are twofold: a) InPars \cite{DBLP:conf/sigir/BonifacioAFN22} utilizes the human-written prompt while SPTAR utilizes an optimized soft prompt; b) SPTAR has a soft prompt filter module to select example document-query pairs. To make it a fair comparison with InPars \cite{DBLP:conf/sigir/BonifacioAFN22}, we choose the same example document-query pairs in the prompt of SPTAR for InPars \cite{DBLP:conf/sigir/BonifacioAFN22} and utilize InPars' original human-written prompt to prompt the LLaMA and Vicuna to obtain weak document-query pairs. We find for InPars' human-written prompt, the quality of generated weak document-query pairs of Vicuna is much better than that of LLaMA, so, for InPars \cite{DBLP:conf/sigir/BonifacioAFN22}, we choose Vicuna as the weak data generator. For SPTAR, LLaMA is better than Vicuna and we choose LLaMA for SPTAR. 

\subsection{Research Questions}
An extensive set of experiments was designed to address the following research questions:

\textbf{RQ1}: Can the proposed SPTAR framework achieve improved performance on DR tasks over the baseline models? (Section~\ref{sec:experi_baseline})

\textbf{RQ2}: During the soft prompt tuning process, does the soft prompt tuning module indeed distill the knowledge from the dataset to the learned soft prompt? What factors contribute to the learned soft prompts? (Section~\ref{sec:effect_of_prompt_tuning})

\textbf{RQ3}: What are the costs of the soft prompt tuning module? Does the soft prompt tuning module greatly increase the training time and computational resources? (Section~\ref{efficient_of_prompt})

\textbf{RQ4}: What specific role does the soft prompt filter play in SPTAR? (Section~\ref{sec: prompt_filter})

\textbf{RQ5}: Can the weak data filter further improve the performances of DR models? (Section \ref{sec: ab_data_filter})

\textbf{RQ6}: For SPTAR's soft prompt tuning module, what is the influence of the size of training data $X$? Is a larger $X$ better than a smaller one? (Section~\ref{sec: train_num_prompt})

\textbf{RQ7}: For SPTAR's soft prompt augmentor module, what is the influence of the number of example document-query pairs $M$? Is a larger $M$ better than a smaller one? (Section \ref{sec: number_example_paris})

% \textbf{RQ7}: What is the influence of different example document-query pairs? Do more example document-query pairs improve the quality of generated weak data?

\begin{table*}[t!]
    \centering
    \resizebox{\linewidth}{!}{%
    \begin{tabular}{cc|cc cc ccc ccc}
      \toprule
        \multicolumn{2}{c|}{Retriever} & \multicolumn{2}{c}{MS MARCO} & \multicolumn{2}{c}{FiQA-2018} & \multicolumn{3}{c}{DL2019} & \multicolumn{3}{c}{DL2020}\\
        \cmidrule(lr){3-4} \cmidrule(lr){5-6} \cmidrule(lr){7-9} \cmidrule(lr){10-12}
        & & MRR@10 & Recall@100 & MRR@10 & Recall@100 & MAP & nDCG@10 & Recall@100 & MAP & nDCG@10 & Recall@100\\
        \midrule
        \multicolumn{2}{c|}{BM25}                & 0.1840 & 0.6578 & 0.2956 & 0.5395 & 0.3013 & 0.5058 & 0.4910 & 0.2856 & 0.4796 & 0.5599 \\
        \multicolumn{2}{c|}{$Contriever_{base}$} & 0.1501 & 0.6412 & 0.1726 & 0.3634 & 0.2187 & 0.4243 & 0.4330 & 0.2324 & 0.4028 & 0.4957 \\
      \multicolumn{2}{c|}{$ReContriever_{base}$} & 0.1606 & 0.6729 & 0.2199 & 0.4871 & 0.2489 & 0.4559 & 0.4806 & 0.2449 & 0.4159 & 0.5177 \\
        % \multicolumn{2}{R|}{TAS-B}               & 0.3039 & 0.8336 & 0.2918 & 0.5166 & 0.3797 & 0.6388 & 0.5554 & 0.3940 & 0.6153 & 0.6430 \\
        \midrule
        \multirow{3}{*}{DPR}       & 
        \multicolumn{1}{|c|}{W/O Aug}            & 0.1303 & 0.5141 & 0.1433 & 0.3738 & 0.1796 & 0.3484 & 0.3289 & 0.2196 & 0.3806 & 0.4218 \\
        &\multicolumn{1}{|c|}{InPars}            & 0.1519 & 0.6092 & 0.2720 & 0.5289 & 0.2474 & 0.4460 & 0.3919 & 0.2407 & 0.3913 & 0.4646 \\
        &\multicolumn{1}{|c|}{SPTAR}             & $\dagger$\underline{0.2114} & $\dagger$\underline{0.7118} & \underline{0.2885} & $\dagger$\underline{0.5747} & $\dagger$\underline{0.3091} & $\dagger$\underline{0.5253} & $\dagger$\underline{0.4928} & $\dagger$\underline{0.3042} & $\dagger$\underline{0.5219} & $\dagger$\underline{0.5585} \\
        \midrule
        \multirow{3}{*}{ColBERT}   & 
        \multicolumn{1}{|c|}{W/O Aug}            & 0.0665 & 0.3157 & 0.1498 & 0.3811 & 0.0679 & 0.2257 & 0.1998 & 0.0654 & 0.1612 & 0.2408 \\
        &\multicolumn{1}{|c|}{InPars}            & 0.1965 & 0.5232 & 0.3031 & 0.4715 & 0.1908 & 0.4618 & \underline{0.3402} & 0.2278 & 0.4645 & 0.4209 \\
        &\multicolumn{1}{|c|}{SPTAR}             & \underline{0.2000} & \underline{0.5312} & $\dagger$\underline{0.3472} & $\dagger$\underline{0.5107} & \underline{0.1923} & \underline{0.4703} & 0.3332 & \underline{0.2338} & \underline{0.4752} & \underline{0.4259} \\
        \midrule
        \multirow{3}{*}{BM25BE}   & 
        \multicolumn{1}{|c|}{W/O Aug}            & 0.1456 & 0.6225 & 0.1659 & 0.4779 & 0.2330 & 0.4211 & 0.4155 & 0.2623 & 0.4236 & 0.5239 \\
        &\multicolumn{1}{|c|}{InPars}            & 0.1660 & 0.6799 & 0.2891 & 0.5700 & 0.2939 & 0.4905 & 0.4720 & 0.2765 & 0.4286 & 0.5552 \\
        &\multicolumn{1}{|c|}{SPTAR}             & $\dagger$\underline{$0.2159$} & $\dagger$\underline{0.7222} & \underline{0.3000} & $\dagger$\underline{0.6046} & $\dagger$\underline{0.3379} & $\dagger$\underline{$0.5596$} & $\dagger$\underline{0.4999} & $\dagger$\underline{0.3194} & $\dagger$\underline{$0.5395$} & $\dagger$\underline{0.5823} \\
        \midrule
        \multirow{3}{*}{BM25CE}    & 
        \multicolumn{1}{|c|}{W/O Aug}            & 0.0876 & 0.5978 & 0.2319 & 0.6013 & 0.2137 & 0.3108 & 0.4337 & 0.1510 & 0.2263 & 0.5055 \\
        &\multicolumn{1}{|c|}{InPars}            & 0.1889 & 0.6230 & 0.3646 & 0.6155 & 0.1549 & 0.2188 & 0.3732 & 0.1425 & 0.1700 & 0.4780 \\
        &\multicolumn{1}{|c|}{SPTAR}             & $\dagger$\underline{0.2009} & $\dagger$\underline{0.7484} & \underline{0.3705} & \underline{0.6193} & $\dagger$\underline{$0.3440^\star$} & $\dagger$\underline{0.5488} & $\dagger$\underline{$0.5459^\star$} & $\dagger$\underline{0.2921} & $\dagger$\underline{0.4405} & $\dagger$\underline{$0.6395^\star$} \\
        \midrule
        \multirow{3}{*}{$Contriever_{be}$}& 
        \multicolumn{1}{|c|}{W/O Aug}            & 0.1872 & 0.7228 & 0.2963 & 0.5920 & 0.2876 & 0.4735 & 0.4959 & 0.2944 & 0.4618 & 0.5851 \\
        &\multicolumn{1}{|c|}{InPars}            & 0.1752 & 0.7253 & 0.3541 & 0.6196 & 0.2704 & 0.4782 & 0.4910 & 0.2748 & 0.4443 & 0.5736 \\
        &\multicolumn{1}{|c|}{SPTAR}             & $\dagger$\underline{0.2148} & $\dagger$\underline{$0.7717^\star$} & $\dagger$\underline{$0.3836^\star$} & $\dagger$\underline{$0.6567$} & $\dagger$\underline{0.3284} & $\dagger$\underline{0.5272} & $\dagger$\underline{0.5402} & $\dagger$\underline{$0.3325$} & $\dagger$\underline{0.5241} & $\dagger$\underline{0.6267} \\
        \midrule
        \multirow{3}{*}{$ReContriever_{be}$}& 
        \multicolumn{1}{|c|}{W/O Aug}            & 0.1938 & 0.7393 & 0.3402 & 0.6212 & 0.2979 & 0.4801 & 0.4913 & 0.2965 & 0.4661 & 0.5902 \\
        &\multicolumn{1}{|c|}{InPars}            & 0.1743 & 0.7367 & 0.3535 & 0.6500 & 0.2914 & 0.4554 & 0.5323 & 0.2849 & 0.4270 & 0.5927 \\
        &\multicolumn{1}{|c|}{SPTAR}             & $\dagger$\underline{0.2121} & $\dagger$\underline{0.7676} & $\dagger$\underline{0.3757} & $\dagger$\underline{$0.6715^\star$} & $\dagger$\underline{0.3343} & $\dagger$\underline{0.5207} & \underline{0.5451} & $\dagger$\underline{0.3258} & $\dagger$\underline{0.5206} & $\dagger$\underline{0.6231} \\
        \midrule
        \multirow{3}{*}{$TAS-B$}& 
        \multicolumn{1}{|c|}{W/O Aug}            & 0.0297 & 0.2046 & 0.1625 & 0.3903 & 0.0525 & 0.1306 & 0.1312 & 0.0388 & 0.0902 & 0.1115 \\
        &\multicolumn{1}{|c|}{InPars}            & 0.2089 & 0.6911 & 0.2582 & 0.5221 & 0.2828 & 0.5257 & 0.4470 & 0.2982 & 0.4812 & 0.5065 \\
        &\multicolumn{1}{|c|}{SPTAR}             & $\dagger$\underline{$0.2541^\star$} & $\dagger$\underline{0.7353} & $\dagger$\underline{0.2943} & $\dagger$\underline{0.5625} & \underline{0.3049} & $\dagger$\underline{$0.5972^\star$} & \underline{0.4737} & $\dagger$\underline{$0.3488^\star$} & $\dagger$\underline{$0.5612^\star$} & $\dagger$\underline{0.5690} \\
      \bottomrule
    \end{tabular}}
    \caption{SPTAR vs baseline models: a) $Contriever_{base}$ is the original Contriever model pre-trained on CC-net and English Wikipedia; b) $ReContriever_{base}$ is the original ReContriever model pre-trained on the same datasets as Contriever; c) W/O Aug doesn't use any augmented data and only applies the training data specified in Table \ref{tab: datasets-partitioin}; d) InPars \cite{DBLP:conf/sigir/BonifacioAFN22} utilizes human-written prompts and it has no soft prompt filter mechanism; e) Within each method, the best results are underscored and the symbols $\dagger$ denote statistically significant enhancements over the second best result, with p-values < 0.05, as determined by a t-test; f) The best results cross different methods are denoted with symbol $\star$. g) Table \ref{tab: datasets-partitioin} documents the data splits for each method, and the filtered example document-query pairs of SPTAR are documented in Appendix \ref{sec: example_pairs_ms} and \ref{sec: example_pairs_fiqa}.}
    \label{tab: main}
    \end{table*}

\section{Experimental Results}

\begin{figure*}[t!]
     \centering
     \begin{subfigure}[b]{0.33\textwidth}
         \centering
         \includegraphics[width=\linewidth]{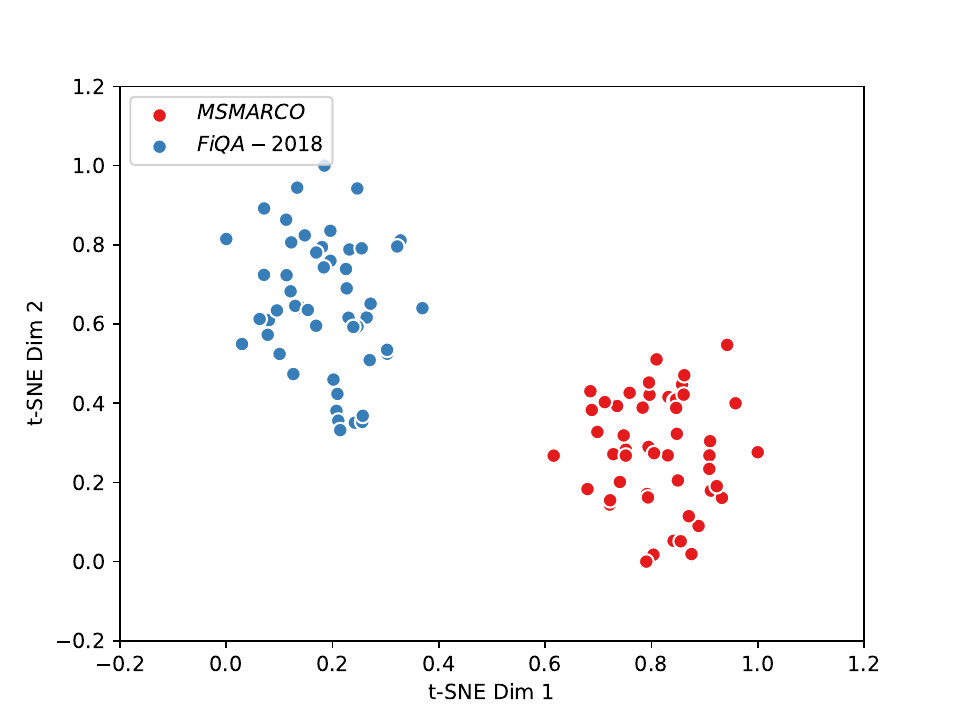}
         \caption{Different datasets.}
         \label{fig:tsne_dataset}
     \end{subfigure}
     % \hfill
     \begin{subfigure}[b]{0.33\textwidth}
         \centering
         \includegraphics[width=\linewidth]{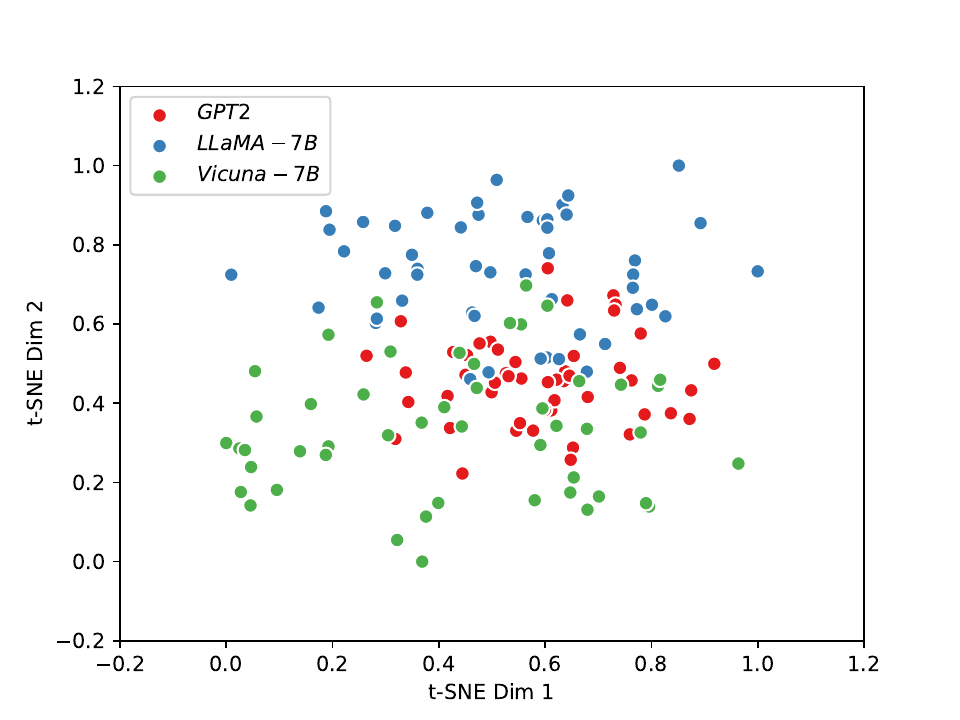}
         \caption{Different LLMs.}
         \label{fig:tsne_llms}
     \end{subfigure}
     % \hfill
     \begin{subfigure}[b]{0.33\textwidth}
         \centering
         \includegraphics[width=\linewidth]{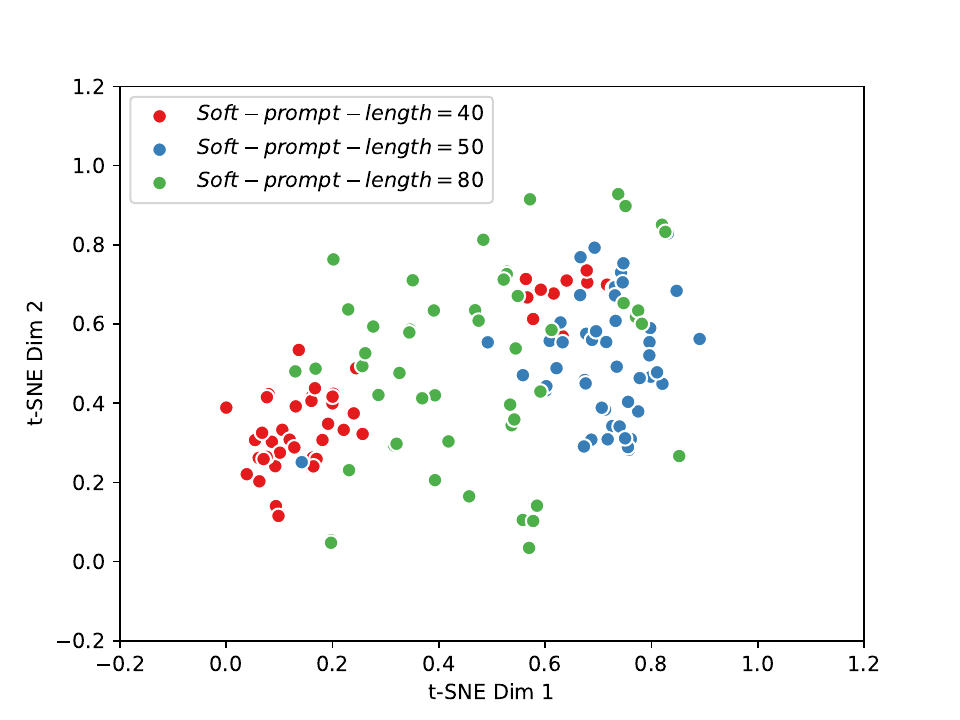}
         \caption{Different lengths.}
         \label{fig:tsne_token_size}
     \end{subfigure}
        \caption{T-SNE embedding visualization of soft prompt's virtual tokens: a) soft prompt’s virtual tokens with different datasets; b) soft prompt’s virtual tokens with different LLMs; c) virtual tokens of soft prompt with different lengths.}
        \label{fig:effect_prompt_tuning}
\end{figure*}

\subsection{SPTAR vs Baseline Models (RQ1)}
\label{sec:experi_baseline}

We conducted evaluations of our SPTAR method on MS MARCO and its related datasets, including DL2019, DL2020, and FiQA 2018. As presented in Table \ref{tab: main}, the SPTAR data augmentation method consistently outperforms established baselines, including W/O Aug and InPars, across a spectrum of widely used datasets and key retrieval metrics only except the R@100 of SPTAR’s ColBERT which is slightly outdone by InPars. TAS-B method gets the most best results (4 out of 10, there are 10 metrics in each row of Table \ref{tab: main}) as it is distilled from two teacher dense retrieval models (cross-encoder and ColBERT) that are both trained on the full MS MARCO dataset. For all 7 retrievers, 55 out of 70 improvements are statistically significant, with p-values < 0.05. ColBERT takes 8 out of 15 un-significant improvements. 

The token-level matching complexity of ColBERT between queries and documents could explain why SPTAR's enhancements on ColBERT did not significantly surpass the InPars baseline on MS MARCO related datasets. DPR’s simple architecture makes it have better generalization ability than ColBERT, which is evidenced by the fact that on MS MARCO related datasets, DPR’ W/O Aug is better than ColBERT’s W/O Aug. Also, there exists noise in the generated queries, especially when the soft prompt is learned from a small dataset (same as the dataset of W/O Aug), and ColBERT’s token-level interaction mechanism is more sensitive than DPR. 
 
Regarding the re-ranking models, BM25CE and BM25BE, both re-rank the top 1000 items initially retrieved by BM25. BM25CE employs a cross-encoder for re-ranking, whereas BM25BE integrates a DPR model. Without data augmentation, BM25BE beats BM25CE on 7 out of 10 metrics when trained on a small labeled dataset (W/O Aug). The cross-encoder model of BM25CE, requiring more extensive data to effectively learn complex interactions and mitigate overfitting risks, ultimately demonstrated superior performance under SPTAR over BM25BE on 6 out of 10 metrics.

The officially released checkpoint of $Contriever_{base}$ is loaded to be evaluated, and we further improved its performance by fine-tuning it as a bi-encoder model denoted as $Contriever_{be}$ in Table \ref{tab: main}. We found performance improvement even fine-tuning $Contriever_{base}$ on a small dataset as Contriever’s W/O Aug is better than $Contriever_{base}$ on all four datasets. Our method, $Contriever_{be}$’s SPTAR, can further improve $Contriever_{base}$ over the second-best value significantly. $Contriever_{be}$’s InPars only beats W/O Aug on 4 out of 10 metrics. Similarly, we evaluated the Re$Contriever_{base}$ and $ReContriever_{be}$ and the same patterns as Contriever are observed as well.

By harnessing the benefits of soft prompt tuning and LLMs, our model generates high-quality weak queries that greatly enhance DR tasks. Moreover, the consistent improvements observed across DPR, TAS-B, BM25BE, BM25CE, Contriever, and ReContriever substantiate the general applicability of our approach, extending beyond specific dense retrievers. It is worth noting that in the absence of augmentation data, all dense retrievers except for the Contriever and ReContriever perform worse than the unsupervised model BM25. This underscores the significant reliance of DR on domain-specific labeled data and highlights the limitations of directly training dense retrievers in scenarios with limited ground-truth data, where the expected performance may not be attainable.

\subsection{Ablation Study}
\label{sec: abstudy}
In this section, our primary objective is to evaluate the distinct contributions of each module to the overall efficacy of the SPTAR framework. Our experiments focus on evaluating the perplexity and NDCG@10 metrics. The perplexity metric, derived from the $S_{eval}^{100}$ dataset, provided insights into the model's text generation quality. The default NDCG@10 scores in this section are obtained by evaluating the SPTAR-DPR model trained, evaluated, and tested on $S_{trail}^{50} + S_{eva}^{100} + W_{small}$, $D_{eval}$ and $D_{test}$ respectively. We didn't filter $W_{small}$ so that the NDCG@10 score can genuinely represent the quality of the weak data.

\subsubsection{The Impact of Soft Prompt Tuning Module (RQ2)}
\label{sec:effect_of_prompt_tuning}
To gain deeper insights into the optimized parameters $\theta^{\ast}$, we employed the t-SNE algorithm \cite{van2008visualizing} to visualize the virtual token vectors of the learned soft prompt $f_{\theta^{\ast}}(s)$ when $\theta^{\ast}$ are converged with different datasets and LLMs. 

Figure \ref{fig:tsne_dataset} illustrates the distribution of virtual token vectors in a two-dimensional space. We utilized the LLaMA-7B language model with a virtual token length $l_{s}=50$ for this experiment. The red and blue points indicate the MS MARCO and FiQA datasets, respectively. The visual analysis clearly reveals that the virtual token vectors from the two datasets exhibit distinct distributions in the two-dimensional space, with minimal overlap. Notably, at the model initialization phase, both datasets share the same prompt $s$, making the observed changes in vector distribution after convergence particularly significant. These findings highlight the remarkable capability of prompt tuning to distill domain-specific knowledge from datasets to the learned prompt token vectors. 
This accomplishment is particularly noteworthy in the scenario where ground-truth data are too limited that human-written prompts struggle to capture domain-specific information and incorporate it effectively into the prompt design.

In Figure \ref{fig:tsne_llms}, various colors distinguish distinct LLMs: GPT-2, LLaMA-7B, and Vicuna-7B. We kept all the hyperparameters the same except for the language model to evaluate the influence of different language models on the parameters $\theta$. The dispersion of points with the same color indicates the extent of parameter updated during training. Figure \ref{fig:tsne_llms} clearly illustrates that the red point cloud representing the GPT-2 model has less dispersion, with points tightly clustered together. In contrast, the blue point cloud representing LLaMA-7B and the green point cloud representing Vicuna-7B exhibit greater dispersion of virtual token vectors. This observation suggests that, when trained on the same dataset, the LLaMA-7B and Vicuna-7B models enable the soft prompt module to absorb more domain-specific knowledge, leading to an enhancement in the generation of synthesized queries. Moreover, similar findings were obtained when decoding the virtual tokens into corresponding words. For instance, after training the GPT-2 model, we observed that the resulting soft prompt merely replicates the prompt tokens used during initialization, essentially duplicating the manual prompt without additional learning. In contrast, when decoding the virtual token vectors into words utilizing the LLaMA-7B and Vicuna-7B, we discovered that these models not only retain the initial prompt tokens but also acquire additional symbols and representations associated with relevant text, such as ``\textit{query}'', ``\textit{rewrite}'', ``\textit{argument}'', ``\textit{enhance}'' and ``\textit{adding}'', indicating parameters $\theta$ does learn task-specific knowledge.

In Figure \ref{fig:tsne_token_size}, we aim to understand the effects of different soft prompt lengths on the tuning module by examining the virtual token vector distribution of the learned soft prompt. This experiment was conducted on LLaMA-7B and dataset MS MARCO and all the hyperparameters are the same except for the soft prompt length. The three lengths 40, 50, and 80 are represented by the colors red, blue, and green, respectively. From the point distribution in Figure \ref{fig:tsne_token_size}, we observe partial overlap between the red and blue points, as well as some distinct points. As the virtual token length increases, the embedding distribution area of the longer soft prompt encompasses the regions corresponding to the shorter ones: 40 and 50. This result aligns with our expectation: with different lengths of soft prompts, the embedding distributions of soft prompts' virtual tokens are different. Nevertheless, regardless of their lengths, the distributions of these prompts tend to have significant overlap and shared regions.

For RQ2, we have conclusions: a) datasets can be distinguished from the learned soft prompts, demonstrating that soft prompt tuning does learn task-specific soft prompts; b) both the LLMs and the length of soft prompts influence the learned soft prompts.

\subsubsection{The Efficiency of Soft-Prompt Tuning (RQ3)}
\label{efficient_of_prompt}

\begin{table}[hpt!]
\centering
\begin{tabular}{ccccc}
\toprule
 LLM  &  $count(\theta)/count(\Phi)$ &  Best Epoch \#\\
\hline
GPT-2 & 0.0308\% & 17 \\
LLaMA-7B & 0.0030\% & 5 \\
Vicuna-7B & 0.0030\% & 4 \\
\bottomrule
\end{tabular}
\caption{Efficiency evaluation of SPTAR's soft prompt tuning module on MS MARCO $S_{train}^{50}$ and $S_{eval}^{100}$ (Section \ref{sec: data_preparation}).}
\label{tab: efficient_soft_prompt}
\end{table}

Table \ref{tab: efficient_soft_prompt} presents the number of learnable parameters and convergence efficiency of soft prompt tuning for different LLMs on the MS MARCO dataset. For the soft prompt tuning module in our proposed SPTAR, despite the vast number of LLM's original parameters $\Phi$, $\Phi$ remains frozen and does not require fine-tuning. The trainable parameters $\theta$ associated with the fine-tuning of the soft prompt are substantially fewer. The percentages in the second column highlight that the soft prompt module's fine-tuning involves an exceptionally small set of parameters $\theta$, roughly equating to \textbf{0.003\%} of the size of $\Phi$. Notably, the size of $\theta$ stays constant, irrespective of the growth of $\Phi$. This characteristic significantly enhances the practicality and training efficiency of SPTAR, as we can fine-tune task-specific soft prompts with a minimal fraction of parameters for optimization.

Furthermore, for a new task or dataset, SPTAR can efficiently complete the fine-tuning process of the soft prompt tuning module within a few epochs. As highlighted in the third column of the table, we examined the convergence speed of the soft prompt tuning model on the evaluation dataset $S_{eval}^{100}$ (Section \ref{sec: data_preparation}) by the best epoch number and the lower this number is, the faster it converges. It becomes apparent that employing a more advanced language model expedites the convergence of the soft prompt tuning module, requiring a mere four or five epochs for convergence. Considering both the count of $\theta$ and the convergence speed, we can confidently conclude that the soft prompt tuning module leverages the advantages offered by LLMs while effectively mitigating the computational resource consumption associated with fine-tuning the whole LLMs.

In conclusion, the soft prompt tuning model only fine-tunes a small part of the parameters $\theta$, and the training converges quickly on LLMs. 
\subsubsection{The Impact of Soft Prompt Filter Module (RQ4)}
\label{sec: prompt_filter}

\begin{table}[hbt!]
\centering
\begin{tabular}{cccccc}
\toprule
Dataset &  Filter &  Perplexity (Dec\%) &  NDCG@10 (Imp\%) \\
\hline
\multirow{2}{*}{MS MARCO} & Worst & 4.1934 & 0.2132\\
 & Best & 3.6649 (+12.60\%) & 0.2376 (+11.44\%)\\
\hline
\multirow{2}{*}{FiQA-2018} & Worst & 410.9207 & 0.1855\\
& Best & 5.7898 (+98.59\%) & 0.1923 (+3.67\%)\\ 
\bottomrule
\end{tabular}
\caption{Evaluation of SPTAR-DPR with the best and worst example document-query pairs in soft prompt augmentor module. SPTAR-DPR is trained on $S_{train}^{X}+S_{eval}^{100}+W_{small}$ and tested on $D_{test}$. Results are obtained on LLaMA-7B. For MS MARCO and FiQA-2018, $M=2$ and $M=1$ respectively.}
\label{tab: prompt_filter}
\end{table}

With the learned parameters $\theta^{\ast}$ in SPTAR's soft prompt tuning module, we observe the example document-query pairs in SPTAR's soft prompt augmentor module do influence the quality of the generated weak data, so it is necessary to select certain $M$ document-query pairs from $S_{train}^{X}$. In this section, we study the impact of SPTAR's soft prompt filter module.
In Table \ref{tab: prompt_filter}, we report the best results of SPTAR-DPR (Section \ref{sec: number_example_paris}): a) for MS MARCO, we report the results of SPTAR-DPR with LLaMA-7B and $M=2$; b) for FiQA-2018, we report the results of SPTAR-DPR with LLaMA-7B and $M=1$. 
The SPTAR-DPR is trained on $S_{train}^{50}+S_{eval}^{100}+W_{small}$ and tested on $D_{test}$. The best and worst $M$ example pairs in Table \ref{tab: prompt_filter} are filtered by the method proposed in Section \ref{sec: soft_prompt_filter}.

As shown in Table \ref{tab: prompt_filter}, the results unequivocally demonstrate that the soft prompt filter significantly enhances performance across all comparisons. Specifically, we observe a noteworthy \textbf{12.60\%} to \textbf{98.59\%} decrease in perplexity and a substantial \textbf{3.67\%} to \textbf{11.44\%} improvement on NDCG@10 in the downstream DPR model. Furthermore, our experimental findings indicate that while the utilization of in-context learning theory, complemented by limited examples, greatly enhances the quality of generated weak queries, the choice of example document-query pairs also exerts a considerable influence on text generation quality. 

\subsubsection{The Impact of Weak Data Filter Module (RQ5)}\label{sec: ab_data_filter}

\begin{figure}[t!]
    \centering
        \begin{tikzpicture}[scale=0.7]
          \begin{groupplot}[group style={group size=1 by 1, vertical sep=35pt}, height=5.88cm, width=9.36cm]
            \nextgroupplot[enlargelimits=0.1, legend style={nodes={scale=0.65, transform shape},at={(0.24,0.99)},
            /tikz/every odd column/.style={yshift=2pt},
            /tikz/nodes={text width=40pt,text depth=,anchor=base},
        }, ylabel={NDCG@10}, xlabel={Top-$k$ of Weak Data Filter}, symbolic x coords={W/O, 10, 30, 50, 70}, xtick=data, enlarge x limits={abs=1cm}, label style={font=\small}, tick label style={font=\small}]
            \addplot coordinates {(W/O,0.2319) (10,0.2558) (30,0.2580) (50,0.2560) (70,0.2557)}; %micro mmoe
            \addplot coordinates {(W/O,0.2242) (10,0.2125) (30,0.2393) (50,0.2318) (70,0.2404)}; %macro mmoe
            \legend{MSMARCO, FiQA-2018}
          \end{groupplot}
        \end{tikzpicture}
        % \vspace{-10pt}
        \caption{SPTAR-DPR NDCG@10 scores with different top-$k$ of weak data filter. SPTAR-DPR is trained on $S_{train}^{50}+S_{eval}^{100}+F_{k}(W_{large})$ (Section \ref{sec: dataset}). Results are obtained on LLaMA-7B. For MS MARCO and FiQA-2018, $M=2$ and $M=1$ respectively.}
    % \Description{}
    \label{fig: ab-weak-data-filter}
\end{figure}
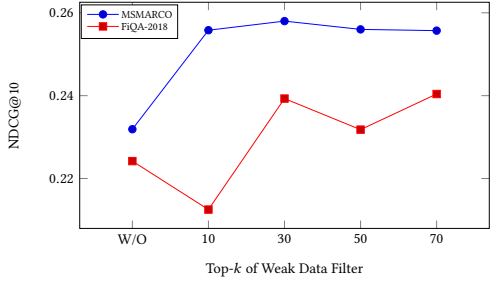

To assess the enhancements achieved by filtering weak data, we applied various top-$k$ values
to filter the generated weak data $W_{large}$, yielding $F_{k}(W_{large})$. We then assessed the
performance of the SPTAR-DPR model, trained on $S_{train}^{50}+S_{eval}^{100}+F_{k}(W_{large})$, on $D_{test}$. This comparison served to quantify the gains over the approach devoid of a weak data filter. Optimal parameters, namely LLM and $M$, were identified and held constant in this section to exclusively assess the influence of top-$k$.

As shown in Figure \ref{fig: ab-weak-data-filter}, on MS MARCO, the SPTAR-DPR model without the data filter gets an NDCG@10 score of \textbf{0.2319} while the score rises to \textbf{0.2580} with data filter top-$k$=30. On FiQA-2018, SPTAR-DPR with filter top-$k$=70 gets the highest NDCG@10 score of \textbf{0.2404}, while it gets an NDCG@10 score of \textbf{0.2242} without a data filter. These consistent gains across different datasets underscore the effectiveness of the weak data filter module (Section \ref{sec: weak_data_filter}). 
We did not discern any correlation between top-$k$ and the NDCG@10 metric; thus, in real-world scenarios, top-$k$ acts as a hyperparameter requiring tuning per dataset.

\subsubsection{The Impact of Training Size $X$ (RQ6)}
\label{sec: train_num_prompt}

\begin{figure}[t!]
    \centering
        \begin{tikzpicture}[scale=0.7]
          \begin{groupplot}[group style={group size=2 by 1, vertical sep=10pt, horizontal sep=40pt}, height=6.0cm, width=6.36cm]
            % \usetikzlibrary{patterns}
            \nextgroupplot[title=Perplexity(Dec\%), title style={yshift=-1ex,}, ybar, bar width=4pt, enlargelimits=0.1, legend style={nodes={scale=0.65, transform shape},at={(0.41, 0.98)},
            /tikz/every odd column/.style={yshift=2pt},
            /tikz/nodes={text width=40pt,text depth=,anchor=base},
        }, ylabel={Percentage}, y label style={at={(0.05,0.5)}}, xlabel={$X$}, symbolic x coords={10, 30, 50}, xtick=data, enlarge x limits={abs=1cm}, label style={font=\small}, tick label style={font=\small}]
            \addplot coordinates {(10, 99.59) (30, 99.74) (50, 99.78)}; 
            % \legend{one-shot, two-shot}
            \nextgroupplot[title=NDCG@10(Imp\%), title style={yshift=-1ex,}, ybar, bar width=4pt, enlargelimits=0.1, legend style={nodes={scale=0.65, transform shape},at={(0.99, 0.98)},
            /tikz/every odd column/.style={yshift=2pt},
            /tikz/nodes={text width=40pt,text depth=,anchor=base},
        }, ylabel={Percentage}, y label style={at={(0.1,0.5)}}, xlabel={$X$}, symbolic x coords={10, 30, 50}, xtick=data, enlarge x limits={abs=1cm}, label style={font=\small}, tick label style={font=\small}]
            \addplot[fill=red!30!white, draw=red] coordinates {(10, 12.17) (30, 27.06) (50, 37.66)}; 
            % \legend{one-shot, two-shot}
          \end{groupplot}
        \end{tikzpicture}
        % \vspace{-10pt}
        \caption{Evaluation of SPTAR-DPR with different $X$ compared with W/O (Section \ref{sec: dataset}). SPTAR-DPR is trained on $S_{train}^{X}+S_{eval}^{100}+W_{small}$ and tested on $D_{test}$. Results are obtained on LLaMA-7B and MS MARCO.}
    % \Description{}
    \label{tab: prompt_training_size}
\end{figure}
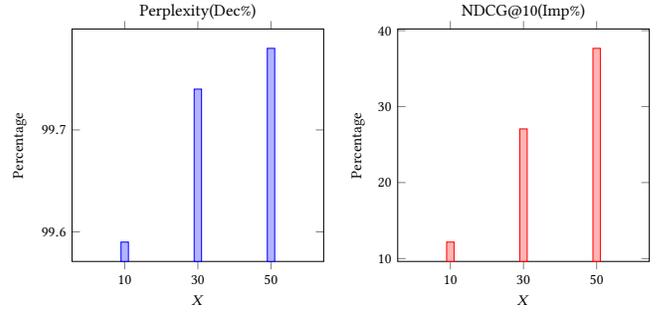

In this section, we analyze the impact of different training sizes $X$ in SPTAR's soft prompt tuning module. To evaluate the impact of $X$, we first conducted soft prompt tuning on $S_{train}^{X}$ and evaluated the perplexity on $S_{eval}^{100}$. Notably, perplexity serves as an intrinsic metric to measure the impact of $X$ on the quality of generated weak queries. Subsequently, we generated $W_{small}$ and tested the SPTAR-DPR model trained on $S_{train}^{X}+S_{eval}^{100}+W_{small}$ on $D_{test}$. NDCG@10 score is applied to measure the impact of $X$ on downstream DR models, like DPR. As shown in Figure \ref{tab: prompt_training_size}, the findings conclusively demonstrate substantial improvements when employing soft prompt tuning with varying training sizes $X$ compared with the results obtained without soft prompt tuning (W/O in Section \ref{sec: dataset}). Specifically, when $X=50$, perplexity is decreased by \textbf{99.78\%}, and an impressive \textbf{37.66\%} enhancement is observed. 
% perplexity is much easier than NDCG@10 to improve, which means that there is a gap between the two metrics.
Interestingly, it’s apparent that enhancing perplexity is more straightforward than improving NDCG@10, suggesting a disparity between these metrics.

Different from InPars \cite{DBLP:conf/sigir/BonifacioAFN22} and Promptagator \cite{dai2022promptagator}, which only utilizes several example document-query pairs in human-written prompts, our findings underscore the benefits of a marginally larger training size $X$ in soft prompt tuning, leading to better performance. This superiority is manifest in the reduced perplexity and the enhanced NDCG@10 scores in downstream tasks with the increment of training size $X$.

\subsubsection{The Impact of Number of Example Pairs $M$ (RQ7)} \label{sec: number_example_paris}

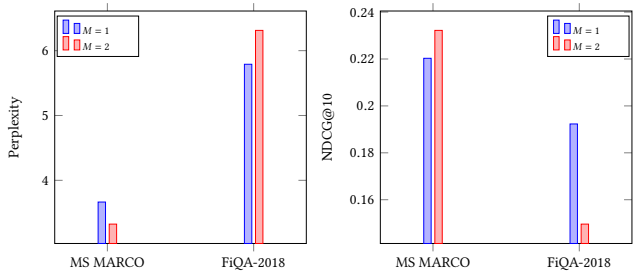
\begin{figure}[t!]
    \centering
        \begin{tikzpicture}[scale=0.7]
          \begin{groupplot}[group style={group size=2 by 1, vertical sep=10pt, horizontal sep=40pt}, height=6.0cm, width=6.36cm]
            \nextgroupplot[ybar, bar width=4pt, enlargelimits=0.1, legend style={nodes={scale=0.65, transform shape},at={(0.335, 0.98)},
            /tikz/every odd column/.style={yshift=2pt},
            /tikz/nodes={text width=40pt,text depth=,anchor=base},
        }, ylabel={Perplexity}, y label style={at={(0.1,0.5)}}, symbolic x coords={MS MARCO, FiQA-2018}, xtick=data, enlarge x limits={abs=1cm}, label style={font=\small}, tick label style={font=\small}]
            \addplot coordinates {(MS MARCO, 3.6650) (FiQA-2018, 5.7901)}; 
            \addplot coordinates {(MS MARCO, 3.3247) (FiQA-2018, 6.3126)}; 
            \legend{$M=1$, $M=2$}
            \nextgroupplot[ybar, bar width=4pt, enlargelimits=0.1, legend style={nodes={scale=0.65, transform shape},at={(0.99, 0.98)},
            /tikz/every odd column/.style={yshift=2pt},
            /tikz/nodes={text width=40pt,text depth=,anchor=base},
        }, ylabel={NDCG@10}, y label style={at={(0.05,0.5)}}, symbolic x coords={MS MARCO, FiQA-2018}, xtick=data, enlarge x limits={abs=1cm}, label style={font=\small}, tick label style={font=\small}]
            \addplot coordinates {(MS MARCO, 0.2203) (FiQA-2018, 0.1923)}; 
            \addplot coordinates {(MS MARCO, 0.2322) (FiQA-2018, 0.1496)}; 
            \legend{$M=1$, $M=2$}
          \end{groupplot}
        \end{tikzpicture}
        % \vspace{-10pt}
        \caption{Evaluation of SPTAR-DPR with different numbers of example pairs $M$. SPTAR-DPR is trained on $S_{train}^{X}+S_{eval}^{100}+W_{small}$ and tested on $D_{test}$. Results are obtained on LLaMA-7B and MS MARCO.}
    % \Description{}
    \label{fig: ab-M}
\end{figure}

In SPTAR's soft prompt agumentor module, when tagging the unlabeled documents with weak queries, $M$ filtered example document-query pairs are utilized to instruct the LLM. In this section, we explore the impact of different $M$. Initially, we selected LLaMA-7B as the LLM and did soft prompt tuning on $S_{train}^{50}$, computing perplexity on $S_{eval}^{100}$. Subsequently, with the filtered $M$ example document-query pairs from SPTAR's soft prompt filter module (Section \ref{sec: soft_prompt_filter}), we generated $W_{small}$. Ultimately, SPTAR-DPR trained on $S_{train}^{50}+S_{eval}^{100}+W_{small}$ is tested on $D_{test}$ to compute NDCG@10. We also did the same experiments on Vicuna, and we found LLaMA-7B model consistently delivers better results than the Vicuna-7B model, no matter whether $M=1$ or $M=2$. Thus, we only report the results on LLaMA-7B in Figure \ref{fig: ab-M}.

As depicted in Figure \ref{fig: ab-M}, for dataset MS MARCO, $M=2$ achieves the best performance in terms of perplexity and NDCG@10. In contrast, for dataset FiQA-2008, $M=1$ demonstrates superior performance. These results contradict our initial assumption that the bigger $M$ is the better the perplexity and NDCG@10 are. We attribute this inconsistency to varying dataset distributions. 
% Given that most QA datasets in which a document has multiple relevant queries and each query is only based on a subset of the document, leading to increased uncertainty and heightened learning complexity for the model. Consequently, these factors contribute to divergent outcomes for different datasets. Thus, we recognize the need for further investigation and exploration of this matter in future studies.
Considering that many QA datasets contain documents with multiple relevant queries, where each query is constructed from a subset of the document, it implies a heightened level of uncertainty and complexity for the learning model. These intricacies, in turn, produce varying performances across different datasets. Therefore, we acknowledge the importance of delving deeper into this topic in subsequent research.

\section{CONCLUSION AND FUTURE WORK}

In this paper, we introduce the soft prompt tuning for augmenting DR (SPTAR) framework to tackle the challenge of limited domain-specific training data in DR tasks. Our approach harnesses soft prompt tuning to optimize soft prompts on limited ground truth data. By prompting LLMs with these optimized soft prompts as well as example document-query pairs, we generate weak queries for unlabeled documents, resulting in an abundant collection of weak document-query pairs for training domain-specific dense retrievers. To further enhance the quality of the generated weak tagged queries, we incorporate a soft prompt filter that selects high-quality example document-query pairs in the prompt as well as a weak data filter module to clean the generated weak data. The effectiveness of our proposed approach is validated through comprehensive experiments. This work represents an initial step toward a promising research direction. In future work, we aim to scrutinize SPTAR’s broad applicability by testing it across diverse datasets. It’s noteworthy that the loss function employed herein is a pointwise loss, implying a suboptimal utilization of negative instances. Future studies might benefit from delving into pairwise and listwise losses. Moreover, there lies potential in probing multi-task soft prompt tuning methods to bolster both efficiency and outcome.

%=========================================================
% \section{Acknoledgments}
% \clearpage
\appendix

\section{Filtered Example Document-query Pairs for MS MARCO and LLaMA} \label{sec: example_pairs_ms}
For MS MARCO, $M=2$ is better than $M=1$ when LLaMA is employed.\\
\subsection{$M=1$}
\textbf{Document}: According to price comparison website Gocompare.com, the cost of becoming a new driver has soared by almost a fifth in the past five years. A survey of 2,000 parents found the average cost of a young driver's first car is Â£3,825, with insurance priced at Â£2,232.Scroll down for video. Expensive: A survey of 2,000 parents found the average cost of a young driver's first car is Â£3,825. The typical learner also needs Â£480 of driving lessons. survey of 2,000 parents found the average cost of a young driver's first car is Â£3,825, with insurance priced at Â£2,232. Scroll down for video. Expensive: A survey of 2,000 parents found the average cost of a young driver's first car is Â£3,825.\\
\textbf{Query}: average insurance cost for new drivers \\
\subsection{$M=2$}
\textbf{Document}: Oakland weather forecast from AccuWeather.com. Extended forecast in Oakland, MD 21550 for up to 25 days includes high temperature, RealFeel and chance of precipitation Oakland weather forecast from AccuWeather.com. Extended forecast in Oakland, MD 21550 for up to 25 days includes high temperature, RealFeel and chance of precipitation my recent locations Â°f Oakland, MD 41Â°\\
\textbf{Query}: weather in oakland md\\
\textbf{Document}: As their name suggests, the triglycerides are composed of one molecule of glycerol and joined via ester bonds with three molecules of fatty acids. As is shown in Figure 12, fatty acids are long chains of carbon and hydrogen usually between 14-24 carbons long (and they always have an even number of carbons).. Phospholipids: This class of lipids are really derivatives of triglycerides. The are composed of a glycerol molecule with two fatty acids (a diglyceride). The third carbon contains a phosphate group and usually some added polar molecule (such as ethanolamine, serine or choline).\\
\textbf{Query}: what are triglycerides composed of \\

\section{Filtered Example Document-query Pairs for FiQA-2018 and LLaMA} \label{sec: example_pairs_fiqa}
For FiQA-2018, $M=1$ is better than $M=2$ when LLaMA is employed.\\
\subsection{$M=1$}
\textbf{Document}: As your is a very specific case, please get an advice of CA. It should not cost you much and make it easier. The sale of agriculture land is taxable in certain conditions and exempt from tax in other cases. Sale of agricultural land is subject to capital gains tax. But there are certain exemptions under Section 54B, subject to conditions, which are as follows: If deemed taxable, you can avail indexation, ie the price at which you grandfather got [the date when he inherited it as per indexation] and pay 10\% on the difference. If the price is not known, you can take the govt prescribed rate. As there is a large deposit in your fathers account, there can be tax queries and need to be answered. Technically there is no tax liable even if your grandfather gifts the money to your father. More details at \url{http://www.telegraphindia.com/1130401/jsp/business/story\_16733007.jsp} and \url{http://www.incometaxindia.gov.in/publications/4\_compute\_your\_capital\_gains/chapter2.asp}\\
\textbf{Query}: Is the amount taxable if my grandfather sells agricultural land \\
\subsection{$M=2$}
\textbf{Document}: Others have already commented on the impact of anything which dissuades merchants from raising possible breaches, so I won\'t dwell on that. Maybe we need stronger legislation, maybe we don\'t, but it doesn\'t change today\'s answer. Often it works the other way around to what you might expect - rather than the merchant noticing and notifying Visa/MC/others, Visa/MC/others spot patterns of suspicious activity (example 1). I don\'t have any data on the relative numbers of who is being notified/notifying between merchants and payment processors, but at the point when your card is identified as compromised there\'s no reason to suppose that an individual merchant in the traditional sense has been compromised, let alone identified. In fact because there\'s a fast moving investigation it could even be a false alarm that led to your card getting cancelled. Conversely it could be a hugely complex multinational investigation which would be jeopardised. It\'s simply not safe to assume that simply ""brand X"" has been compromised, therefore everything ""brand X"" knows about you is also compromised: Furthermore there\'s no reason to assume the merchant has even admitted to, or discovered the root cause. MC/Visa/Banks, at the point at which they\'re cancelling cards simply can\'t say (at least not in a way that might expensively backfire involving lots of lawyers) because the standard of proof needed to go on record blaming someone is simply not yet met. So: yes it\'s common that you aren\'t told anything for all of the above reasons. And of course if you really want to find out more you may have some success with your local data protection legislation and formally make a subject access request (or local equivalent) to see what that brings back. Be sure to do it in writing, to the official address of both mastercard and your bank.\\
\textbf{Query}: MasterCard won't disclose who leaked my credit card details\\
\textbf{Document}: Surprised nobody has mentioned Freshbooks yet. It's lightweight, easy to use, and free for low-end use (scaling up price-wise as you scale up).\\
\textbf{Query}: What's the best application, software or tool that can be used to track time?\\
% \clearpage
\bibliographystyle{ACM-Reference-Format}
\balance
\bibliography{references}

% \clearpage
% \appendix
% \section{Reproducability}

\end{document}